\documentclass[trackchanges]{aastex701}

\newcommand{\Wm}{W\,m$^{-2}$}

\newcommand{\So}{So}
\newcommand{\Snom}{${S}^{\rm N}_{\odot}$}

\newcommand{\Lnom}{\hbox{$\mathcal{L}^{\rm N}_{\odot}$}}

\newcommand{\PSUAA}{Department of Astronomy \& Astrophysics, 525 Davey Laboratory, 251 Pollock Road, Penn State, University Park, PA, 16802, USA}
\newcommand{\PSUCEHW}{Center for Exoplanets and Habitable Worlds, 525 Davey Laboratory, 251 Pollock Road, Penn State, University Park, PA, 16802, USA}
\newcommand{\PSETI}{Penn State Extraterrestrial Intelligence Center, 525 Davey Laboratory, 251 Pollock Road, Penn State, University Park, PA, 16802, USA}
\newcommand{\nasagoddard}{NASA Goddard Space Flight Center, Greenbelt, Maryland, USA}

\shorttitle{Solirads}
\shortauthors{Mamajek et al.}

\begin{document}

\title{The {\it solirad} (So) as a Convenient Unit for Quoting\\
Astronomical Irradiances for Planetary Insolations and Exoplanetary Instellations}

\author[0000-0003-2008-1488]{Eric E. Mamajek}
\affiliation{Jet Propulsion Laboratory, California Institute of Technology, M/S 321-162, 4800 Oak Grove Drive, Pasadena, CA 91109, USA}
\email[show]{mamajek@jpl.nasa.gov}
\correspondingauthor{Eric E. Mamajek}

\author[0000-0001-6160-5888]{Jason T.\ Wright}
\affiliation{\PSUAA}
\affiliation{\PSUCEHW}
\affiliation{\PSETI}
\email{astrowright@gmail.com}

\author[0000-0003-3989-5545]{Noah W. Tuchow}
\affiliation{Steward Observatory and Department of Astronomy, The University of Arizona, Tucson, AZ 85721, USA}
\affiliation{\nasagoddard}
\email{nwtuchow@arizona.edu}

\author[0000-0003-1705-5991]{Patrick A. Young}
\affiliation{School of Earth and Space Exploration, Arizona State University, 781 E Terrace Mall, Tempe, AZ 85287, USA}
\email{patrick.young.1@asu.edu}

\author[0000-0002-7064-8270]{Matthew A. Kenworthy}
\affiliation{Leiden Observatory, Leiden University, Einsteinweg 55, NL-2333 CC Leiden, The Netherlands}
\email{kenworthy@strw.leidenuniv.nl}

\author[0000-0002-0388-8004]{Emily A. Gilbert}
\affiliation{NASA Exoplanet Science Institute-Caltech/IPAC, 1200 East California Boulevard, Pasadena, CA 91125, USA}
\email{egilbert@ipac.caltech.edu}

\begin{abstract}
Measurements of physical parameters for stars and (exo)planets are often quoted in units normalized to the Sun and/or Earth. The {\it nominal total solar irradiance}, ${S}^{\rm N}_{\odot}$, while based on a current best estimate with uncertainties, was adopted to be an exact reference value of 1361 W\,m$^{-2}$ by IAU 2015 Resolution B3, corresponding to ``{\it the mean total electromagnetic energy from the Sun, integrated over all wavelengths, incident per unit area per unit time at distance 1 au}''. In the planetary and exoplanetary science literature, the units employed for ``flux'', ``insolation'', ``instellation'', etc., are often cumbersome or inconsistent. To simplify the quoting of irradiance units for astronomical applications, we introduce the portmanteau {\it solirad}, short for {\it {\bf sol}ar {\bf i}r{\bf rad}iance}, as an abbreviated version of the longer IAU term ``nominal total solar irradiance''. The solirad (So) is a unit of irradiance, where 1 {\it solirad} = 1 So = 1361 W\,m$^{-2}$, equivalent to the IAU nominal total solar irradiance, and to an apparent bolometric magnitude of $m_\mathrm{bol} \simeq -26.832$\,mag (per IAU 2015 Resolution B2). 
\end{abstract}


\section{Introduction \label{sec:intro}}

The motivation for this paper is to propose a 
convenient shorthand term for quoting units for irradiances in various astronomical subfields (for e.g., (exo)planetary science, astrophysics, heliophysics).
The proposed unit -- the {\it solirad} (So) -- is meant as a shorthand term to complement, and not replace, the SI unit of irradiance (W\,m$^{-2}$).
The unit may also be useful for comparing heat (or thermal) fluxes in other physical contexts (e.g., tidal, geothermal, biological).\\

\subsection{The Need for a Simple Astronomical Irradiance Unit \label{sec:need}}

The total --- or bolometric\footnote{In the context of astronomical observations, the apparent bolometric flux $f_{bol}$ is the total flux, or power, received from an object outside the atmosphere, where $f_{bol}$ = $\int_{0}^{\infty} f_{\lambda}\,{\rm d}\lambda$ = $\int_{0}^{\infty} f_{\nu}\,{\rm d}\nu$, where $f_{\lambda}$ and $f_{\nu}$ are the {\it spectral irradiance} or {\it flux density} \citep[e.g.,][]{Bessell1998,Rieke2003,Gray2008}.} --- amount of electromagnetic energy per unit time (or power) incident upon a plane per unit area is called the {\it irradiance} or {\it heat flux density} \citep[e.g.,][]{Wilkins1990,Rieke2003,SIUnits}.
When describing incident solar radiation upon solar system bodies, on a plane normal to the Sun -- the irradiance is sometimes called {\it incident solar flux} or {\it flux} \citep[e.g.,][]{Lissauer2013}, which may be refer to a time-averaged value \citep[e.g.,][]{Pierrehumbert2010}. 
Besides {\it insolation}, when describing stellar radiation 
incident upon exoplanets, numerous terms are used in the literature:
{\it incident stellar flux} \citep[e.g.,][]{Gaidos2004,delGenio2019}, 
{\it stellar flux} \citep[e.g.,][]{Kopparapu2020}, 
{\it stellar effective flux} \citep[e.g.,][]{Graham2020}, 
{\it stellar irradiance} \citep[e.g.,][]{Meadows2010}, 
{\it stellar insolation} \citep[e.g.,][]{Gaudi2005}, 
{\it insolation flux} \citep[e.g.,][]{Batalha2014}, 
{\it instellation} \citep[e.g.,][]{DomagelGoldman2011,Bryson2021}.
The values may be either reflecting an instantaneous irradiance evaluated at some orbital radius, $r$; an irradiance evaluated at the semi-major axis, $a$; an orbit-averaged value taking into account the eccentricity \citep[e.g.,][]{Eastman2019}; or a ``{\it surface instellation}'' of flux reaching the planet's surface \citep[e.g.,][]{Lobo2024}. 
Regardless, the terms are consistent with units of energy per time per area, or W\,m$^{-2}$ in SI.  
The heliophysics community measures {\it total solar irradiance} (TSI), evaluated at the standard distance\footnote{In 2012, IAU General Assembly adopted the astronomical unit to be precisely 1\,au = 149597870700\,m; see IAU 2012 Resolution B2 ``{\it on the re-definition of the astronomical unit of length}'' at \href{https://www.iau.org/Iau/Iau/Science/Resolutions.aspx}{https://www.iau.org/Iau/Iau/Science/Resolutions.aspx}. 
This effectively decoupled the {\it au} from either the Earth's time-varying orbit, or the combination of the Gaussian gravitational constant $k$ \citep{Gauss1809,Clemence1965} and estimates of the solar mass parameter $\mu_{\odot} = G M_{\odot}$ \citep[see ][]{Klioner2008}}.
of 1\,au \citep[e.g.,][]{Steinhilber2009,Kopp2011,Solanki2013}. 
It was also historically called the {\it solar constant} \citep[e.g.,][]{Eddy1982,Kuhn1988}, a ``constant'' whose variations could be reliably measured by the late 20th century.
Regardless of which irradiance term is adopted, whether it refers to an instantaneous or orbit-averaged value, the Sun or another star, etc., {\it how} the value is quoted has shown a variety of descriptions and units. 
We review the current state of references to irradiance, demonstrate the need for a simpler moniker, and propose an alternative.\\

\section{Towards a Convenient Irradiance Unit \label{sec:towards}} 

The energy received by a planet from its host star (or stars) is one of the most important factors impacting its characterization, and critical for modeling and interpreting planetary observational data. 
Indeed, the stellar energy that a planet receives is often the dominant contributor to the energy budget of a planet's atmosphere and surface, and in the case of a terrestrial planet, governs its climate and surface conditions. 
The concept of the habitable zone --- the region around a star where a planet could host surface liquid water --- is parameterized in terms of the fluxes that planets receive from their host star \citep[e.g.,][]{Kasting1993}. 
For stars with sufficiently characterized masses and ages, the history of incident flux on a planet's atmosphere can be predicted from stellar evolutionary tracks, and can serve as one of the key drivers of planetary climatic evolution and the atmospheric escape. 
The history of incident stellar flux can be used to understand the long-term habitability of exoplanets, examining the duration that planets spend in the habitable zone and the continuity of habitable conditions throughout their evolution \citep{Tuchow2024,Ware2025}. 
Continuously habitable planets, characterized via their flux history, may be the ideal candidates for searches for biosignatures, and incident stellar fluxes need to be sufficiently constrained in order to characterize the atmospheric compositions of observed exoplanets \citep{Gaudi2020}.\\

Here, we show some examples of how such values are quoted for exoplanets. It is clear that a more concise unit would ease communication of this straightforward concept.\\

In the literature on hot Jupiter exoplanets, one occasionally encounters units of ``flux'' (irradiance/insolation) quoted in units of Gerg\,s$^{-1}$\,cm$^{-2}$ \citep[e.g.,][]{Fortney2021,Kawashima2021,Ma2021}.
Quoting in SI (W\,m$^{-2}$) would reduce the units from 3 to 2, and remove 2 prefixes. 
The unit {\it Gerg s$^{-1}$ cm$^{-2}$} is neither SI, nor a ``simple'' non-SI cgs unit (requiring two prefixes), nor even a convenient astronomical constant (i.e., normalized to that of a standard measurement or parameter for a commonly studied celestial body).\\

The NASA Exoplanet Archive\footnote{\url{https://exoplanetarchive.ipac.caltech.edu/}} \citep{Akeson2013,Christiansen2025} quotes ``{\it insolation flux}" as ``{\it S (S$_{\Earth}$)}'' in units of ``{\it Earth flux}''.
However, the ``{\it Earth flux}'' could be construed as 
the {\it intrinsic} heat flux of the Earth itself, which has a surface-averaged value of $\sim$92 mW\,m$^{-2}$, and which may be considered the ``Earth flux'' from a geophysics perspective \citep[][]{Davies2010}.\\

The description of the quantity and how to quote its units is illustrated by the \citet{Kopparapu2014} study of the limits of the habitable zone as a function of planet and stellar mass.
The study alternatively quotes the quantity in units of 
``{\it effective solar flux ($S_\mathrm{eff}$)}'' and 
``{\it effective flux incident on the planet ($S/S_\mathrm{o}$)}'',
corresponding to ``{\it Earth's flux}'' and ``{\it effective stellar flux}''.\\

The effective flux, $S_\mathrm{eff}$, is used extensively in the exoplanet literature, especially in the context of planetary habitability and climate of rocky planets. 
It occupies a similar conceptual space to the irradiance, making it easy to conflate the two when coming from a background with different terminology. 
It is, however, a fundamentally different parameter. 
$S_\mathrm{eff}$ introduces additional complications because it is not a fixed quantity related to the star's total luminosity, but rather it is defined by energy balance between incoming flux and outgoing IR flux at the surface of the atmosphere.
It depends on radiative transfer in the atmosphere and various other planetary processes such as atmospheric energy transport, absorption and re-emission by the planetary surface, and energy produced by the planet itself. 
Radiation transport alone dictates that the value of $S_\mathrm{eff}$ will depend upon the spectrum of the incident stellar flux -- and therefore the star's $T_\mathrm{eff}$ -- as well as the assumed atmosphere of the planet.
\citet{Kopparapu2020} quotes the ``{\it instellations}'' and ``{\it stellar incident fluxes}'' in ``{\it Earth units}'' and ``{\it starlight on planet relative to sunlight on Earth}''.\\

Several other examples of these units can be found in the literature.
The transit and precision radial velocity exoplanet literature employ multiple terms and choices of units.
The ExoFAST code \citep{Eastman2013,Eastman2019} quotes the orbit-averaged ``{\it incident flux}"\footnote{\citet{Eastman2019} defines the orbit-averaged flux incident to the planet to be 
$\langle$$F$$\rangle$ = $\sigma_B T^4_{\rm eff} \{ \frac{R_*}{a} (1 - e^2 / 2) \}^2 $, where $\sigma_B$ is the Stefan-Boltzmann constant, $T_{\rm eff}$ is the stellar effective temperature, $R_*$ is the stellar radius, $a$ is the orbital semi-major axis, $e$ is the orbital eccentricity.} $\langle$$F$$\rangle$ in units of $10^9$ erg\,s$^{-1}$\,cm$^{-2}$.
The Kepler Data Processing handbook \citep{Jenkins2020} calculated the ``{\it planet effective stellar flux $\phi_\mathrm{eff}$ defined as the ratio of the flux of the host star at the top of planet's atmosphere to the solar flux at the top of Earth’s atmosphere}''.
The TESS validations reports have quoted ``{\it effective stellar fluxes}'' in units of ``{\it Goldilocks}''.
A recent paper on the discovery of the exoplanet GJ~12b shows typical language: \citet{Kuzuhara2024} states ``{\it [T]he insolation flux that Gliese 12 b receives from the host star is calculated to be $1.62^{+0.13}_{-0.11}$ $S_{\Earth}$, where $S_{\Earth}$ is the insolation of the Earth.}" 
\citet{Lichtenberg2024} states ``{\it The latter planets are, with a few exceptions, all highly irradiated to $\gtrsim$ 30 Earth’s instellation at present day ($F_\mathit{Earth}$).}"
\citet{delGenio2019} states ``{\it ...incident stellar flux from 0.6 to 1.2$\times$ Earth’s solar constant S$_o$}'' with data in a table column labeled ``{\it incident stellar flux relative to Earth}''.  
Clearly there is a wide range of how insolations/instellations/fluxes and their units are quoted in the astronomical literature.\\

\subsection{Choice of a Standard Irradiance \label{sec:choice}}

Two IAU 2015 Resolutions - B2 and B3\footnote{\href{https://www.iau.org/Iau/Publications/List-of-Resolutions}{https://www.iau.org/Iau/Iau/Science/Resolutions.aspx}, see also \citet{Mamajek2015B2} and \citet{Mamajek2015B3}.} - adopted a nominal total solar irradiance, \Snom, equivalent to exactly 1361\,\Wm. When \Snom\, was combined with the IAU 2012 definition of the astronomical unit, it provided a {\it nominal solar luminosity} (\Lnom) value of 3.828$\times10^{26}$\,W. 
The choice of adopted nominal total solar irradiance value was reviewed by \citet{Prsa2016}, and it represents a rounded value representative of the recent total solar irradiance values from spaceborne experiments accounting for averaging over a solar cycle \citep[e.g.,][]{Kopp2011}. 
The value adopted for the IAU resolution (1361 \Wm) was estimated to be accurate to $\pm1$\,\Wm (2$\sigma$) accounting for absolute calibration uncertainty $\sim$0.03\% \citep{Kopp2014}.
Subsequent analyses in the heliophysics literature remain consistent with this value \citep[e.g.,][]{Lean2020,Kopp2024,Richards2024,Kopp2025}.\\

\subsection{Choice of Name for Irradiance Unit \label{sec:name}}

While the IAU {\it nominal total solar irradiance} is an accurate description of the constant, we propose a more succinct term.
Several candidate abbreviated names for the irradiance unit were considered and then dismissed:\\  

\begin{itemize}
\item {\it Sol} would be duplicative with the unit of solar days for Mars, and is not specific enough given the number of other normalized solar parameters.
\item {\it Soli} has the benefit of being short and included parts of the term nominal total {\bf sol}ar {\bf i}rradiance.
However, the term has numerous other uses in various fields: in music, law, the name of a people and language, and the term is used by multiple businesses and part of many trademarks. 
\item {\it Langley} is a previously defined non-SI, cgs unit of solar heat transmission, equal to 1 cal\,cm$^{-2}$ (41.840 kJ\,m$^{-2}$) \citep{Gyllenbok2018}. 
While recognizing an early pioneer of measuring the solar irradiance, the name was already used, for a different type of unit (energy per unit area), and defined such that it does not represent any useful quantity normalized to that measured at Earth (either on the ground or in space) with respect to the total solar irradiance \citep[Earth receives approximately 2 Ly s$^{-1}$;][]{Gyllenbok2018}.
\item {\it Solar Constant} is an old term, which unfortunately 
usually refers to measurements of the {\it time-varying} solar irradiance itself, hence the ``constant'' changes over time \citep[e.g.,][]{Eddy1982,Kuhn1988,Frohlich1995}.
The term is also somewhat ambiguous as to what ``constant'' 
parameter is being referenced. 
Given the historical baggage of the term, it was considered best to avoid it.
\end{itemize}

We adopt the portmanteau {\bf solirad} as it contains the key parts of the term total {\it sol}ar {\it i}r{\it rad}iance, in a very abbreviated form.
As of December 2025, the word {\it solirad} had no entry in either the Oxford English or Merriam-Webster\footnote{\url{https://www.oed.com/} and \url{https://www.merriam-webster.com/}} online dictionaries.
For the abbreviation, we've adopted a capitalized version - So\footnote{As opposed to $S_0$ or $S_o$.}.
SI units and related ``non-SI units accepted for use with the SI units'' are typically only capitalized if they are based on a proper name (e.g., ``K'' for Kelvin), with the exception of {\it Liter}\footnote{
\url{https://www.bipm.org/en/committees/cg/cgpm/16-1979/resolution-6}}.
Unfortunately adopting a lower case abbreviation ({\it so}) might be problematic given the myriad uses of ``so'' in the English language\footnote{\url{https://www.oed.com/search/dictionary/?scope=Entries&q=so}}.
\So\, is also somewhat similar to previous symbols sometimes found in the literature (e.g., $S_{\odot}$, $S_{\Earth}$).\\

\subsection{Discussion and Examples of Usage \label{sec:usage}}

Here we provide some examples of use of the term {\it solirad}, along with SI prefixes for this non-SI term. 
The ``conservative'' habitable zone range from \citet{Kopparapu2014} is 0.356 to 1.107 solirads (\So), while the ``optimistic'' range covering the cases of ``Recent Venus'' and ``Early Mars'' covers irradiances of 0.32 to 1.776 \So.
The minimum threshold stellar incident flux for which radius inflation is observed among ``hot Jupiter'' exoplanets is estimated to be $\sim2\times10^{8}$ erg\,s$^{-1}$\,cm$^{-2}$ = 2 MW\,m$^{-2}$, equivalent to 1470 \So\, (1.47 k\So).
The irradiances for the solar system planets range from approximately $\sim$6.7 \So\, for Mercury to $\sim$0.0011 \So\, (1.1\,m\So) for Neptune.
Averaged over the 25 year span 2000-2025, the JPL Horizon's system\footnote{\url{https://ssd.jpl.nasa.gov/horizons/}} shows that the osculating orbital elements for the Earth-Moon barycenter around the Sun has average eccentricity $e$ = 0.0167025 and semi-major axis $a$ = 149597937.7 km = 1.00000044816712 au. Assuming the nominal total solar irradiance at 1\,au, the orbit-averaged solar flux is then $<F>$ = $(a / 1\,{\rm au})^{-2} (1 - e^2)^{-1/2}$ \So\, \citep{Williams2002}, or about 1.00013 solirad.\\ 

The term could also be used for heat fluxes.
Io's average observed heat flux due to tidal heating is $\sim$2.25 W\,m$^{-2}$ \citep{Tyler2015}, or about 0.00166 \So\, (1.66 m\So). 
The Earth's internal heat flux \cite[$47\pm2$ TW;][]{Davies2010} averaged over its surface is $\sim$0.092 W\,m$^{-2}$, or about 0.000068 \So\ (68 $\mu$\So).
The skin of human adults has an average heat flux of $\sim$10 mW\,cm$^{-2}$ = 100 W\,m$^{-2}$ \citep{Thielen2018}, or about $\sim$0.07 \So.\\

Some conversions for common irradiance units to solirads are listed in Table \ref{tab:units}.\\

\begin{deluxetable}{lll}[htb!]
\tablecaption{Conversions for Common Flux Units}
\label{tab:units}
\setlength{\tabcolsep}{4pt}
\tablehead{
\colhead{Unit} & \colhead{in SI} & \colhead{in So}
}
\startdata
1 W\,m$^{-2}$ & 1 W\,m$^{-2}$ & $7.34754\times10^{-4}$ \So\\
1 kW\,m$^{-2}$ & 10$^3$ W\,m$^{-2}$ & 0.734754 \So\\
1 mW\,m$^{-2}$ & 10$^{-3}$ W\,m$^{-2}$ & $7.34754\times10^{-7}$ \So\\
1 erg\,s$^{-1}$\,cm$^{-2}$  & 10$^{-3}$ W\,m$^{-2}$ & $7.34754\times10^{-7}$ \So\\
1 Gerg\,s$^{-1}$\,cm$^{-2}$ & 10$^{6}$ W\,m$^{-2}$ & 734.754 \So\\
\enddata
\end{deluxetable}

Fluxes in units of \So\, can also be relayed to an {\it apparent bolometric magnitude} (per IAU 2015 Resolution B2):  

\begin{equation}
m_\mathrm{bol} = -2.5\,{\rm log} ( f / f_{\circ}) = -2.5\,\log f - 18.997351...\\
\end{equation}

\noindent where $f$ is flux or irradiance in SI units (W\,m$^{-2}$), $f_{\circ}$ = 2.518021002$\times$10$^{-8}$ W\,m$^{-2}$ is the IAU zero point of the apparent bolometric scale (defining $m_\mathrm{bol}$ = 0). 
For $f$ = 1\,\So\, = 1361\,\Wm, the apparent bolometric magnitude is $m_\mathrm{bol} \simeq -26.832$\,mag.

\begin{equation}
m_\mathrm{bol} = -2.5\,{\rm log}\,( f / \mathrm{So}) = -2.5\,{\rm log}\,f_\mathrm{So} - 26.832
\end{equation}

\noindent which for distance $d$ = 1\,au\, translates into absolute bolometric magnitude (for the nominal Sun) of $M_\mathrm{bol,\odot}$ = 4.74.\\

\section{Conclusion \label{sec:conclusion}}

Occasionally, new terms or units are helpful to simplify communication for astronomical research, e.g., astronomical unit, parsec, etc. 
We propose that in astronomical calculations one can quote the related parameters ``insolation'', ``instellation'', ``total solar irradiance'', ``effective solar flux'', ``stellar flux'', ``heat flux density'', etc., in units of {\it solirads}, where 1 {\it solirad} (\So) equals exactly 1361 W\,m$^{-2}$ -- equivalent to the IAU 2015 nominal total solar irradiance.

\begin{acknowledgments}
The research was carried out in part at the Jet Propulsion Laboratory, California Institute of Technology, under a contract with the National Aeronautics and Space Administration (80NM0018D0004).
We thank Jessie Christiansen, Stephen Kane, and Ravi Kopparapu for their comments.    
\end{acknowledgments}

\bibliography{ms}{}
\bibliographystyle{aasjournalv7}

\end{document}